# A Prony method variant which surpasses the Adaptive LMS filter in the precision of the output signal's representation of the input

Parthasarathy Srinivasan

BeeHive Software Solutions, USA

parth.vasan@gmail.com

*Abstract :* The Prony method for approximating signals comprising sinusoidal/exponential components is known through the pioneering work of Prony in his seminal dissertation in the year 1795. However, the Prony method saw the light of real world application only upon the advent of the computational era, which made feasible the extensive numerical intricacies and labor which the method demands inherently. The Adaptive LMS Filter which has been the most pervasive method for signal filtration and approximation since its inception in 1965 does not provide a consistently assured level of highly precise results as the extended experiment in this work proves. As a remedy this study improvises upon the Prony method by observing that a better (more precise) computational approximation can be obtained under the premise that adjustment can be made for computational error, in the autoregressive model setup in the initial step of the Prony computation itself. This adjustment is in proportion to the deviation of the coefficients in the same autoregressive model. The results obtained by this improvisation live up to the expectations of obtaining consistency and higher value in the precision of the output (recovered signal) approximations as shown in this current work and as compared with the results obtained using the Adaptive LMS Filter.

## 1. Introduction

The Prony method is an effective observation on the transformation of an exponential expression to a mathematically convenient and tractable polynomial form.

The cornerstone of the method is to setup the autoregressive model where the subsequent values of the input signal (are assumed to) depend regressively on the prior values and then solve the model (by dividing the Toeplitz matrix with the input signal and obtaining the remainder) to obtain the autoregression coefficients(a).

The characteristic polynomial summing the autoregression coefficients(a) is then solved to obtain the roots. The roots obtained directly provide the value of the damping factor and frequency of the output approximating signal. The roots obtained are then also, used

(in the form of knowns) to again setup and solve the original autoregressive model to obtain the Amplitude, and phase, of the approximating (or smoothened for lack of a better term) output signal. As obvious the procedure described above is mathematically(computationally) intensive and precise only up to the precision allowed for in the computational process.

Thus, the Prony method was practicable only with modern computers came to aide. This made the computing task easier however introduced a limit (albeit smaller than manual computation) on the precision to which the resultant (approximating) signal resembled the original input signal.

In this current work, the author makes the empirical premise that the computer introduces an imprecision proportional to the variance of the signals' autoregression coefficients and therefore can be thus corrected to yield a more precise output estimate of the Amplitude, frequency, dampener, and phase.

The Adaptive LMS filter method can be described in simple terms as a technique in which the filter coefficients(weights) are adjusted to converge to (theoretical) optimal values where the filter approximates to the desired characteristics. The value of the weight adjustment is obtained by computing the difference between the desired output and the Adaptive Filter output at each iteration and taking a Mean Square of the difference as the quantum of weight adjustment in multiplicative conjunction with the constant convergence multiplier, whose value is chosen somewhat arbitrarily following a suitable procedure described later in this work. Experimentation with this setup of the Adaptive LMS Filter and evaluation of the results using the Precision Measure (again defined later in this work) , show unequivocally that the results obtained for precision do not provide any guarantee or assurance on the lower bound for precise signal representation of the output signal viz. a viz. the input signal. This is in sharp contrast to the adaptation of the Prony Method found as part of this work, which does provide that assured lower bound on the precision making this a superior alternative candidate.

## 2. Context of Current Work in regard to Recent Literature of Related Works

In the recent (century) time frame (~2014,2022) there have been various attempts to comprehend, adapt, implement, and optimize the Prony method with modern electronic and computing platforms doing the heavy duty tasks. These adaptations take the form of varying the parameters in the core Prony Algorithm (with optimization in perspective)**[7]**, generating usable intermediary output byproduct from the core algorithm (for use in related settings and contexts)**[5]**, evaluating the utility of the transformative power (in the co-domain) of the algorithm for ease of scientific analysis**[6]** etc. Also, as described above the Adaptive LMS Filter has been employed albeit there is a lacuna as proven experimentally by this work. However, the objective of this present work is more subtle and profound, in that the adaptation proposed here-in has wide-reaching consequences (in terms of results) in critical applications across diverse domains and application functional areas.

## 3. Scope and significance of Work

This work is focused on obtaining reliable bounds in the computational precision of the output from the Prony method. The bounds so obtained are significant in the practical application of the method owing to the requirement that, for the diverse domains in which the method is applied, such as biomedical engineering, power systems etc. the level of precision significantly impacts the ends for which the method is used , for example the monetary cost effectiveness in the case of power systems and the effectiveness of diagnosis in the case of bio-medical engineering. The work is unique in that though works with similar objectives exist , the particular premise and methodology described are not found in existing literature and this work surpasses the existing methods , including the Adaptive LMS Filter method.

## 4. Fundamental premise in this adaptation of Prony's method

The implementation of the Prony's method in MATLAB is described in **[2]**. As mentioned above, in this adaptation of the Prony method, the coefficients of the characteristic polynomial, computed in the first step autoregressive model are self-adjusted by a constant multiple of the standard deviation of their own values and this assuredly compensates for the imprecision in the computation due to the computing device's limitation in capacity to perform the computation to an arbitrary precision.

## 5. Methodology Employed in this adaptation of Prony's method

**Step 1 :**

This step consists of setting up an Auto Regressive Model with the random sample observations, where the assumption is that the observations are naturally organized such that the nth observation is linearly dependent on the n-1 pre-ceding observations thus yielding a model which regressively relies on computing the Toeplitz matrix set up to relate those observations in accordance with the above assumption.

**Step 2 :**

Solving the above auto-regressive model yields coefficients of the Prony characteristic polynomial which is pre-requisite to obtain the Amplitude and Frequency of the final approximating signal output as yielded by the polynomial roots.

**Step 3 : (Nuanced Prony method proposed by this work)**

Adjust the coefficients of the characteristic polynomial by a standard deviation of their own values. This work proves that this adjustment helps overcome the computational imprecision introduced due to limitations of the computing devices' arbitrary precision arithmetic.

**Step 4 :** Work with the adjusted polynomial coefficients to obtain the Amplitude and Frequency of the output signal components and also compute the damping factor and phase by yet another derivative system of equations obtained by substitution of obtained roots in the original equation.

**Step 5 :** Compute the Precision Metric defined and observe that the nuance proposed here-in lives up to the expectation of yielding constant lower bound in the precision of the output signal approximation.

## 6. The Comparative, Adaptive LMS Experiment

In order to compare the Adaptation of the Prony Method found in this work, the following Adaptive LMS filter was setup in MATLAB(SIMULINK) (based upon the standard example provided by the Simulink software MATLAB/SIMULINKexample: dsp\ViewTheCoefficientsOfYourAdaptiveFilterExample) :

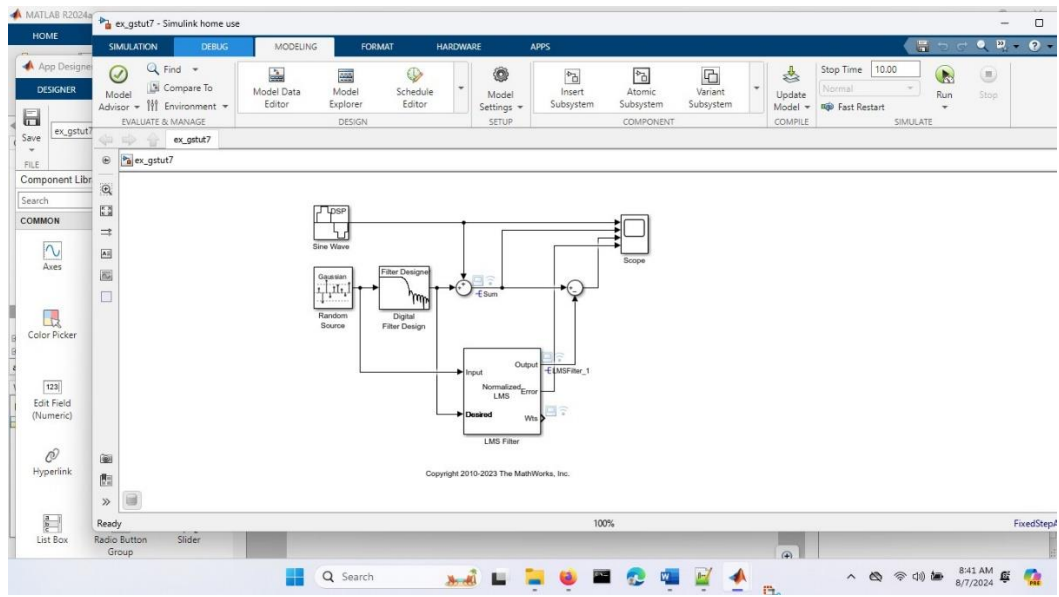

Fig 6(a) : Shows the Model setup for the Adaptive LMS Filter (Refer Introduction Section [1])

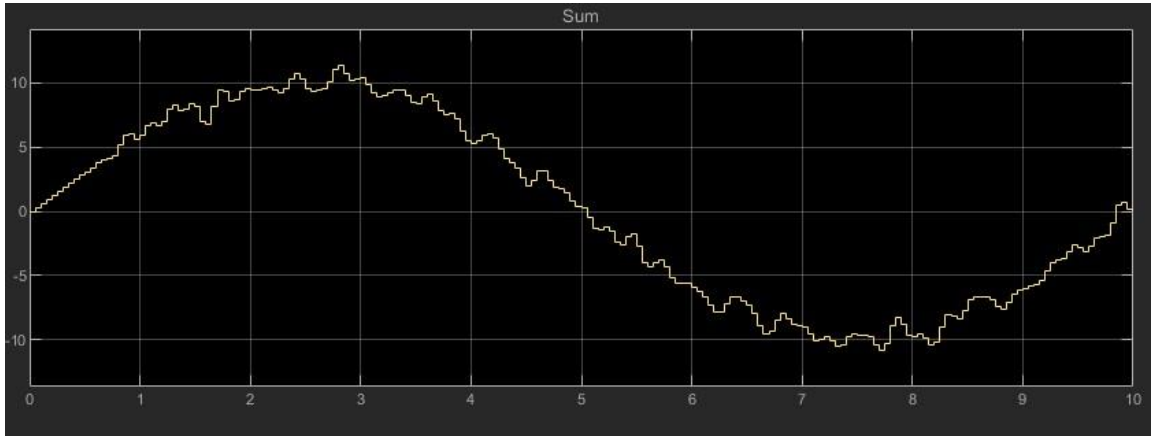

Fig 6(b) : Shows the Input Sine wave "Summed" with the Gaussian Random Noise (The Input Signal)

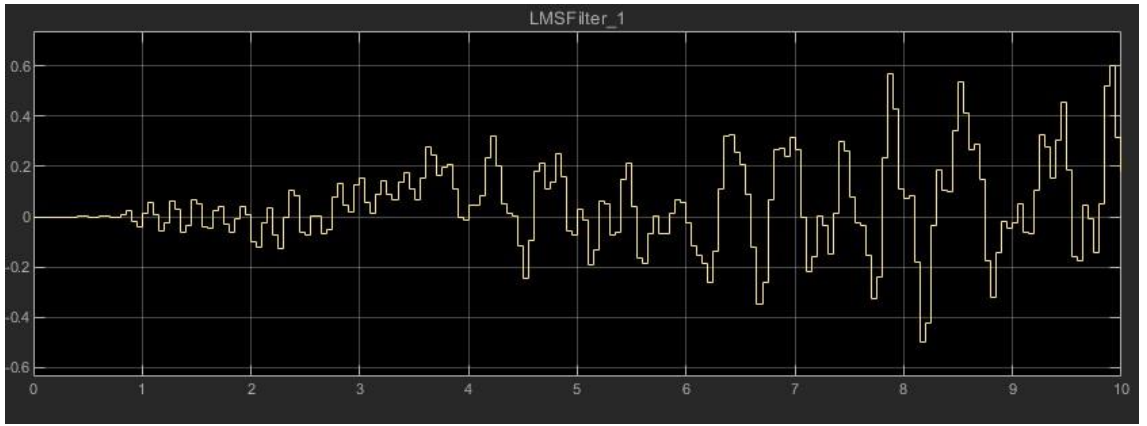

Fig. 6 (c) : Shows the Output Signal from the Adaptive LMS Filter (The Output Representative of the Input Signal)

7. **Following Initialization parameters were used for the Adaptive LMS Filter setup**

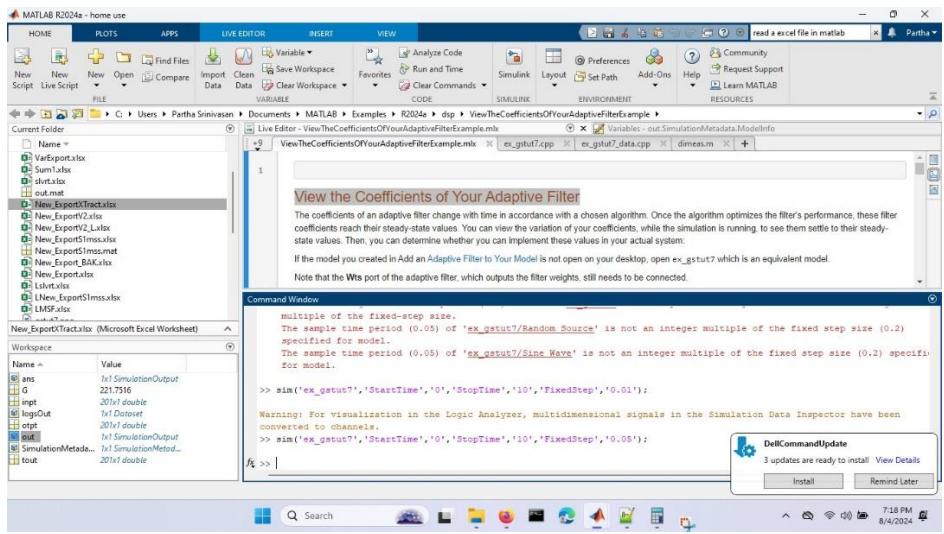

Fig. 7 : Initialization of Adaptive LMS.

## 8. Precision Measure obtained for the Adaptive LMS

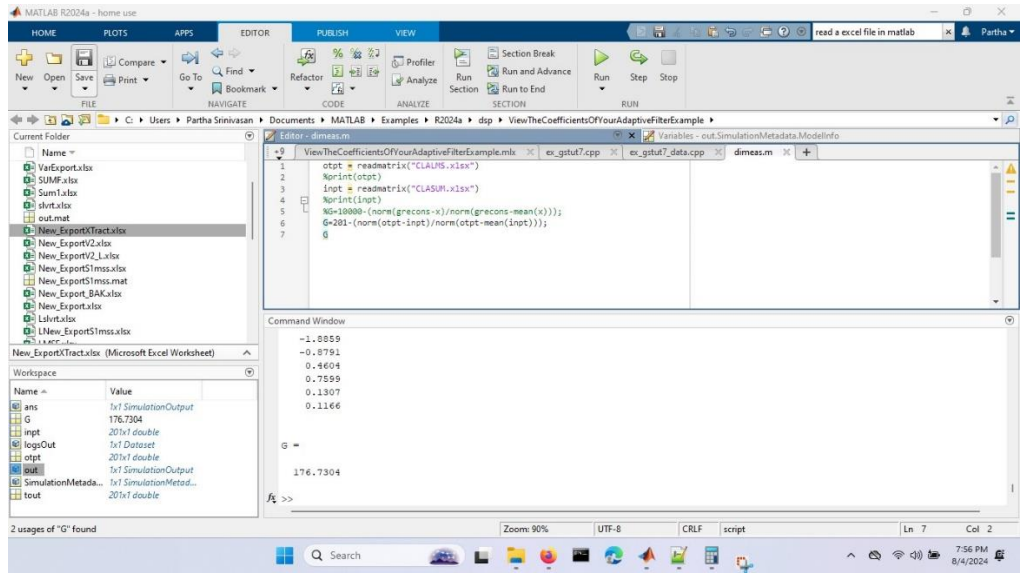

## 9. Advantages of the Prony Adaptation Found here compared to the existing Adaptive LMS method

1) The Adaptive LMS Filter is Heavily dependent on the hugely arbitrary choice of convergence coefficient (mu) for signals with low and high pass characteristics which are increasingly constituting a large chunk of signals now appearing in several practical applications.

2) Even with a proper choice of (mu) the precision of signal recovery(output viz. a viz. input) has been found to vary hugely without any assurance/guarantee on the recovery precision with the Adaptive LMS Filter method.

3) The New Prony Adaptation Approach found here completely eliminates any need for arbitrary choices for parameters for (weight) adjustments , since the coefficients are now adjusted by their own (known) standard deviation with a known multiplier.

4) A guaranteed assurance on the lower bound of the precision is obtained with the Adapted Prony method, as proven by experimentation as against the Adaptive LMS method which does not yield that precision guarantee across varying parameters, as proven by experiment.

5) Also, the Prony Method Adaptation is computationally lightweight
with a one step adjustment to the coefficients as opposed to the
Adaptive LMS Filter which requires coefficient computation and
adjustment at each iteration.

## 10. Details of hardware and software used in the experimental emulation

| Hardware | Operating System | Software Library |
|---|---|---|
| PC 64 bit (AMD64) | Windows 10 | MATLAB R2024a |

## 11. Definition of Metric to quantify the precision of the recovered output signal

To quantify the improvement in precision with this adaptation to the Prony method the following formulation (similar to **[1]**) is used as a measure of the comparative similarity in the input and output signals in both the unchanged Prony method and the adaptation introduced in this work.

Precision Measure,

PM = N – (norm(grecons[i]-g[i])/norm(grecons[i]-mean(g[i]))),

where N=size(g[i]), g[i] = input signal vector, grecons[i] = reconstructed output signal vector, norm is the standard 2 norm,
and mean is the usual mean definition.

## 12. Experimental Results of the Precision Measure for this Prony Method Adaptation.

The results are divided into 3 distinct tiers based on the size of the input signal set vector, to the computation, as follows :

| Input Vector Size | Coefficient Adjustment Factor | Precision Measure Value (consistently for this adaptation of Prony method) (constant across several program runs) |
|---|---|---|
| 100 | 1*sigma(coefficients) | 90 |
| 1000 | 10*sigma(coefficients) | 968.3772 |
| 10000 | 100*sigma(coefficients) | 9900 |

The intuitive rigor behind the empirical constant results obtained above by experiment seems to be the fact that the amount memory required for the computer's mathematical operations in terms of precision grows linearly with the variance of the values being computed itself. This is likely an aggregate manifestation of the optimizations of the memory architecture implementation.

## 13. Experimental Results of the Precision Measure for the LMS Filter

| Input Vector Size | Precision Measure Value (consistently for this adaptation of Prony method) (constant across several program runs) |
|---|---|
| 225 | 174.79 |
| 202 | 193.35 |

## 14. Conclusion and Future Work

This work has proven that with the adaptation of the Prony method set forth by the Author there is a consistent improvement of the Lower Bound in the recovered signal precision, thus improvising the Prony Method for advanced use in diverse and sensitive fields such as Bio-Medical engineering and Electrical Power Systems.

Also , by experimentation described above the Prony Method Adaptation has been proved to surpass the existing Adaptive LMS filter method in the consistent Bound obtained for the Precision of the Output Signal with respect to the Input.

The Bio-Medical applications include improvement to disease detection by signal analysis as in [1].

There is scope for rigorous Stability Analysis (of recovered signal) with the adjustment founded by this work in place.

Also, there is scope to apply this work in the realm of Quantum Computing to evaluate whether the same empirical results are upheld in that realm as well.

# Appendix A -→ MATLAB Code of modified Prony Method

```
%Code 1 Polynomial Implementation
function [Amp,alfa,freq,theta]=polynomial_method (x,p,Ts,method)
% method: 'classic', 'ls' or 'tls' (case insensitive)
% define the solving methods
CLASSIC = 0;
LS = 1;
TLS = 2;

N = length(x);

if strcmpi(method,'classic')
if N ~= 2*P
disp ('ERROR: length of x must be 2*p samples in classical method.');
Amp = [];
alfa = [];
freq = [];
theta = [];
return;
end
else solve_method = CLASSIC;
end

%% step 1

T = toeplitz(x(p:N-1) ,x(p:-1:1));

switch solve_method
case {CLASSIC, LS}
a = -T\x(p+1:N);

case TLS

a= tls(T,-x(p+1:N));

end

%% step 2
c = transpose([1; a]);
r = roots(c);

rprim = r - (std(r)/n);
%rprim = r - std(r); %TBTD
```

```matlab
alfa = log(abs(r))/(2*pi*Ts);

% In case alfa equals to +/-Inf the signal will not be recovered for n=0
%(Inf*0 = Nan). Making alfa = +/-realmax that indeterminance will be solved
alfa(isinf(alfa))=realmax*sign(alfa(isinf(alfa)));

%% step3
switch solve_method
    case CLASSIC
        len_vandermonde = p; % exact case (N=2p) find h with p samples
    case LS
        len_vandermonde = N; % overdetermined case (N>2p) find h with N samples
    case TLS
        len_vandermonde =N; % overdetermined case (N>2p) find h with N samples
end

Z = zeros(len_vandermonde,p);
for i=1:lenght(r)
    
    z(:,i) = transpose(r(i).^(0:len_vandermonde-1));
end

rZ = real(Z);
iZ = imag (Z);
% here Inf values are substituted by realmax values
rZ(isinf(rZ))=realmax*sign(rZ(isinf(rZ)));
iZ(isinf(iZ))=realmax*sign(iZ(isinf(iZ)));

z = rZ+1i*iZ

switch solve_method
    case {CLASSIC,LS}
        h = Z\x(1:len_vandermonde);
    case TLS
        % if exists nan values SVD won't work
        indeterminate_form = sum(sum(isnan(Z) | isinf(Z)));
        if (indeterminate_form)
            Amp = []; alfa = []; freq = []; theta = [];
            return;
        else
            h = tls(Z,x(1:len_vandermonde))
        end
end
Amp = abs(h);
theta = atan2(imag(h),real(h));
```